\documentclass[preprint2]{aastex}
\usepackage{natbib}
\shorttitle{Small Dust in Protoplanetary Disks}
\shortauthors{Kelling and Wurm}

\begin{document}

\title{{\normalsize{\normalfont{Published in \textit{The Astrophysical Journal}, 733:120 (5pp), 2011}}}\\A Mechanism to Produce the Small Dust Observed in Protoplanetary Disks}


\author{T. Kelling and G. Wurm}
\affil{Fakult\"at f\"ur Physik, Universit\"{a}t Duisburg-Essen, Lotharstrasse 1, 47057 Duisburg, Germany}
\email{thorben.kelling@uni-due.de}

\begin{abstract}
Small (sub)-micron dust is present over the entire lifetime of protoplanetary disks. As aggregation readily depletes small particles, one explanation might be that dust is continuously generated by larger bodies in the midplane and transported to the surface of the disks. In general, in a first step of this scenario, the larger bodies have to be destroyed again and different mechanisms exist with the potential to accomplish this. Possible destructive mechanisms are fragmentation in collisions, erosion by gas drag or light induced erosion. In laboratory experiments we find that the latter, light induced erosion by Knudsen compression and photophoresis, can provide small particles. It might be a preferred candidate as the dust is released into a low particle density region. The working principle of this mechanism prevents or decreases the likelihood for instant re-accretion or re-growth of large dense aggregates. Provided that there is a particle lift, e.g. turbulence, these particles might readily reach the surface of the disk.
\end{abstract}





\section{Introduction}
It has become common knowledge that protoplanetary disks exist for a time span of up to 10 million years. The best evidence for this comes from observations of an infrared excess, of which the probability of detection varies with the age of star formation regions (\cite{haisch2001}). These observations are sensitive for small particles. A more detailed modeling of the spectral energy distribution of such disks and emission features at 10 $\mu$m show that, at least in part, this is due to the existence of small particles of the order of 1 $\mu$m (\cite{olofsson2010}). The existence of even smaller 100 nm particles typical for interstellar conditions has also been reported (\cite{meeus2001}, \cite{acke2004}). The existence of small grains at the surface of protoplanetary disks for millions of years is not trivial to explain. Aggregation by relative motion, collisions, and sticking rapidly depletes the small grains and leads to the formation of larger aggregates (\cite{dullemond2005}). Eventually these aggregates get compacted and rain out to the midplane. To solve the missing small particle problem two alternatives are possible for preventing the agglomeration and sedimentation at a certain stage. Charging of dust particles might be one possible explanation as suggested by \cite{okuzumi2009}. The second alternative might be the destruction of larger bodies and the feedback of the smaller particles from the midplane to the surface. Along this line of reasoning a few candidates exist that could destroy the large bodies.

The first mechanism is fragmentation by collisions. At higher collision velocities, collisions can result in small dust particles (\cite{wurm2005}, \cite{teiser2009}, \cite{schraepler2011}) and turbulence might transport these dust particles back to the surface. There are two caveats with this mechanism. First, collisions will produce the dust in a cascade that usually results in a power-law size distribution (\cite{dohnany1969}, \cite{mathis1977}). This way. producing small dust particles implies that mass is also present in the larger mass ranges. Such a continuous size distribution might not be consistent with the total mass available in the disk; i.e. T-Tauri disks have a large mass fraction in millimeter-size particles and adding additional mass bins of similar mass might amount to more mass than is available in the system. Second, the dust will be embedded in a dense environment and will readily collide with other particles again. On the way to the surface collisions will lead to re-growth and eventually compact aggregates will again rain out. Wheter a fraction of dust could survive its way back to the disk without forming aggregates is subject to future simulations and is not yet settled.

The second mechanism is erosion by gas drag in the simple sense as deserts are eroded by wind on Earth. This effect has been studied in laboratory experiments by \cite{paraskov2006} who showed that indeed erosion by gas drag is possible in dense disks for slightly eccentric dusty bodies. This certainly will occur but to what degree is uncertain. One advantage is that the size distribution is not continuous but the larger body only produces small aggregates. This might be more consistent with the total mass budget of a disk. However, it shares the problem of re-accretion or re-growth of particles on the way back to the surface. It is also unclear if the mechanism can provide micron-size grains.

Evaporation of inward drifting material and re-condensation is also a way to provide small particles but this would imply that the dust will primarily consist of high temperature components and crystalline which is not what is observed (\cite{olofsson2010}). Instead of these well-known mechanisms as a possible candidate for dust generation, we consider  light-induced erosion mechanisms.

In \cite{wurm2006}, \cite{wurm2007}, \cite{wurm2008}, \cite{kocifaj2010} and \cite{kelling2011} we showed that the illumination of a dusty body in a low ambient pressure environment leads to eruptions of dust as outlined below. Especially at the inner edge of a disk this might disassemble dusty bodies (\cite{wurm2007}). With respect to explaining dust at the surface of protoplanetary disks this mechanism has a few features that make it a prime candidate for explaining dust at the surface of disks. It provides dust in a low-density environment, and large grains instantly get separated from the small grains. Therefore, small grains can remain small over a long time period and small particles can be transported to the disk surface. The mechanism does not change the size distribution throughout the disk as it only erodes the small fraction of larger bodies at the inner edge; therefore, it will not interfere with the overall mass budget of the disk. Also, as there is evidence that high-temperature minerals that formed close to the star are transported to the outer regions of protoplanetary disks as, e.g., found by the \textit{Stardust} mission (\cite{zolensky2006}), the inner disk is already known as a reservoir of dust particles.

In this paper, we describe the basics of the light-induced erosion mechanisms. We report on laboratory experiments, which show that indeed small grains are generated as part of the erosion process and show that typically smaller grains are separated from larger grains in a way that keeps the small particles at low particle density.

\section{Photophoretic eruptions, Knudsen compression, and photophoretic particle sorting}
In laboratory experiments \cite{wurm2006} found that a light absorbing dust bed, which is illuminated with optical radiation at a flux larger than a few kW m$^{-2}$ eruptively loses particles. \cite{kelling2011} recently showed that a dust bed also loses, for a certain time span, even more particles if the light source is turned off. In protoplanetary disks this might be the case close to the terminator or for moving shadows from surface unevenness of an illuminated rotating body.

Both erosion processes are related to temperature gradients, which are established in the top layers of the illuminated dusty body. This has been modeled in detail by \cite{kocifaj2010,kocifaj2011}. In the illuminated case, radiation can penetrate to a depth some $\mu$m below the surface but thermal radiation cools the surface right at the top. This could be described as a solid state greenhouse effect (\cite{niederdoerfer1933}, \cite{kaufmann2007}). As a consequence, a temperature gradient evolves, where the maximum temperature is below the surface and not at the top.

In the case where the illumination is turned off, the surface cools by radiation. This also leads to a temperature gradient where the surface is cooler than the dust below the surface. The temperature gradient (cool at the surface, warm below) in this case is established over a much larger depth (millimeters). Forces on the particles, which lead to dust eruptions, are generated if the dust bed is situated in a gaseous environment. It is known for more than a century that temperature gradients lead to gas flows along particle surfaces called thermal creep (\cite{knudsen09}) and that interaction with gas molecules at different sides of a particle with different temperatures directly leads to a momentum transfer to the particle called photophoresis (\cite{rohatschek1995}).

The effects of thermal creep (\cite{hettner1924}) and photophoresis (\cite{beresnev1993}) are non trivial. For special cases (large Knudsen numbers, mean free path of gas molecules is much larger than the relevant scale), a rather intuitive description can be given.
We first note that thermal creep refers to the motion of the gas while photophoresis refers to the motion of a solid. Both occur in gas / solid systems where temperature gradients exist. Both are related to each other by momentum conservation.
Nevertheless, both might best be visualized by different models. 

\textit{Thermal creep}. Consider two gas reservoirs 1 and 2, with different temperatures $T_1$ and $T_2$ and different pressures $p_1$ and $p_2$ that are connected by a tube of diameter $s$, which is much smaller than the mean free path of the gas molecules $\lambda$. In equilibrium, the gas flow rates from both sides through the tube must equal $n_1v_1=n_2v_2$. As $n\propto p$ and $v\propto \sqrt{T}$, it follows that the pressures in the chambers are related by $p_2/p_1 = \sqrt{T_2/T_1}$ (\cite{knudsen09}). Hence, in the warmer chamber an overpressure $\Delta p$ is established. For $Kn \simeq 1$ the overpressure is directly proportional to the temperature difference $\Delta p \propto \Delta T$ (\cite{muntz2002}). Simply speaking (\cite{hettner1924}), a surface element of the connection between the chambers is impacted by gas molecules with a greater momentum from the hotter side of the tube. Hence, the tube is affected by a tangential force in the direction from warm to cold. Due to momentum conservation, the tube itself applies a tangential force to the gas in the opposite direction and hence the gas moves from cold to warm.

\begin{figure}
\epsscale{.80}
\plotone{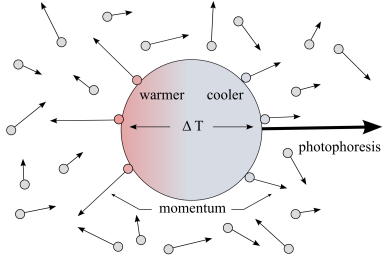}
\caption{Photophoresis. Gas molecules (small circles) with thermal momentum accommodate to the surface of a suspended particle (large circle). They take over the local surface temperature and leave with a momentum according to that temperature -- molecules from the warmer side of the particle leave with a larger momentum than the molecules from the cooler side. Hence, a force (photophoresis) accelerates the particle in the direction from warm to cold.}
\label{fig:photopho_principle}
\end{figure}

\textit{Photophoresis}: The motion of a solid particle is caused by an interaction of the surface of a suspended particle and the surrounding gas molecules (Figure \ref{fig:photopho_principle}).
In principle, gas molecules with thermal momentum accommodate to the particle's surface, take over the local surface temperature, and leave the surface again with a momentum according to their new temperature (\cite{rohatschek1995}). If a particle has a temperature gradient over its surface, the local surface temperature varies. Gas molecules accommodating to the warmer side leave the surface  therefore with a larger momentum than the gas molecules accommodating to the cooler parts of the surface. As a result, a force acts on the suspended particle accelerating it in the direction from warm to cold. The temperature gradient over the particle might be established by illumination, hence the name photophoresis.  For perfectly absorbing, spherical particles at large Knudsen numbers, the photophoretic force is given as (\cite{rohatschek1995})

\begin{equation}
F_{ph_{\Delta T}} = \frac{\pi \alpha p}{T} \frac{\Delta T}{6 }a^2\label{eq:photopho1},
\end{equation}

where $\alpha$ is the accommodation coefficient of the particles surface, $p$ is the ambient pressure, $a$ is the particle radius, $T$ is the mean temperature, and $\Delta T$ is the temperature difference over the particle surface. 

\begin{figure}
\epsscale{.80}
\plotone{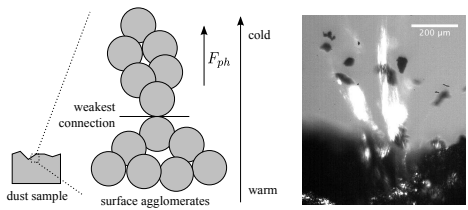}
\caption{Photophoretic ejections (illumination is present). The left figure shows the principle of the photophoretic ejections (from \cite{kelling2011}) and the right figure is a microscopic image of basalt ejections illuminated with $\sim 50$ kW m$^{-2}$ red laser. If the strength of the induced photophoretic force overcomes cohesion st the weakest connection and gravity, a surface agglomerate is ejected.}
\label{fig:photopho}
\end{figure}

\begin{figure}
\epsscale{.80}
\plotone{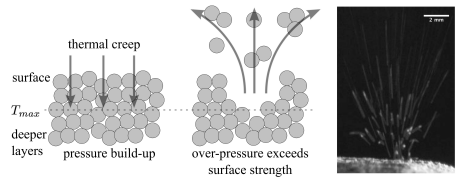}
\caption{Knudsen compressor ejections (illumination is switched off). On the left, the basic principle is depicted and the right image shows an actual ejection spot of a basalt sample shortly after the illumination was switched off (from \cite{kelling2011}). Due to the extended sub surface temperature maximum after the light is switched off, a Knudsen compressor effect (also thermal creep) leads to an overpressure below the surface. If this overpressure exceeds the local tensile strength and gravity particles are ejected.}
\label{fig:knudsen}
\end{figure}

Photophoresis and the effects of thermal creep can lead to particle ejection.
Depending on the temperature difference, which depends on the illuminating flux and dust bed parameters, dust particles at the surface can directly be ejected by photophoresis (see also Figure \ref{fig:photopho}).
If the temperature gradient spans deeper into the dust bed (covers more layers of dust, e.g., after the light source is switched off), thermal creep can lead to a gas pressure increase below the surface analogous to a compressor built by \cite{knudsen09}. If the pressure difference is large enough, dust is ejected from the dust bed (Figure \ref{fig:knudsen}). Then the ejection is not a direct response to thermal creep like photophoresis but a slower build-up of pressure by means of thermal creep. \cite{kelling2009} demonstrated that single dust agglomerates can be levitated over a hot surface by such a Knudsen compressor effect.

Once particles have been released from a dusty body in a protoplanetary disk, an already existing slow gas flow will transport the particles away from their parent body. Therefore a cloud of small particles then moves within the protoplanetary disk which is still illuminated by the central star. Photophoresis acts on the particles but this time it is not generated by a temperature gradient at the top of a dust bed but by direct illumination with stellar light. For a particle with constant thermal conductivity $\kappa_p$ Equation (\ref{eq:photopho2}) can be written as (\cite{rohatschek1995})

\begin{equation}
F_{ph_I} = \frac{\pi \alpha p}{T}\frac{J_1 I}{3\kappa_p}a^3.\label{eq:photopho2}
\end{equation}

Here, $J_1$ denotes the so-called asymmetry factor (heat source distribution function within the illuminated particle) which is set to $\left| J_1\right| = 1/2$ for an opaque sphere, $I$ is the incident light flux and $\kappa_p$ is the particles thermal conductivity. The photophoretic force $F_{ph_I}\propto a^3$. Photophoresis will accelerate the particles to a velocity where the gas drag equals the photophoretic force. Gas drag at large Knudsen numbers can be written as

\begin{equation}
F_{gas} = \frac{mv}{\tau}\label{eq:gasdrag},
\end{equation}

where $m$ is the particle mass, $v$ is the particles velocity and $\tau$ is the gas-grain coupling time which can be expressed as (\cite{blum1996})

\begin{equation}
\tau = \epsilon \frac{m}{\sigma}\frac{1}{\rho_g v_m}\label{eq:tau},
\end{equation}

with an empirical factor $\epsilon$, $m$ as the particles mass, $\sigma$ as the geometrical cross section of the particle, $\rho_g$ as the gas density, and $v_m = \sqrt{(8k_B T)/(\pi m_g)}$ the mean thermal velocity of the gas molecules with $k_B = 1.38\times 10^{-23}$ J K$^{-1}$ as the Boltzmann constant and $m_g = 3.9 \times 10^{-27}$ kg as the molecular mass of the gas molecules . Hence, Equations (\ref{eq:gasdrag}) and (\ref{eq:tau}) show that the gas drag is $F_{gas} \propto a^2$ (with $\sigma \propto a^2$). In equilibrium the drift velocity away from the star therefore linearly depends on the particle size ($v\propto a$) with 

\begin{equation}
v = \frac{\alpha \epsilon J_1 p I}{3 \kappa_p \rho_g T v_m}a\label{eq:vdrift1}.
\end{equation}

With $p=\rho_g T R/M$, where $M = 2.34\times 10^{-3}$ kg mol$^{-1}$ is the molar mass of the gas and $R = 8.3$ J mol$^{-1}$ K$^{-1}$ is the universal gas constant, one finds for the drift velocity
\begin{equation}
v = \frac{\alpha \epsilon J_1 R I}{3\kappa_p M}\frac{1}{\sqrt{(8k_B T)/(\pi m)}}a\label{eq:vdrift2}.
\end{equation}
Typical values $\alpha = 1$, $\epsilon = 0.7$, $J_1=0.5$, $I = 10$ kW m$^{-2}$ at $0.1$ AU, $\kappa_p = 0.01$ W m$^{-1}$K$^{-1}$ and $T= 10^3$ K (\cite{wood2000}) yield $v\simeq 10^{-1}$ m s$^{-1}$ for a 1 $\mu$m particle and $v\simeq 10$ m s$^{-1}$ for a 100 $\mu$m particle. Larger grains are therefore rapidly pushed outward where -- eventually -- they are pushed into an optically dense region and likely are reprocessed in one way or the other, e.g., taking part in reaggregation.

However, the small (sub) micron fraction stays in the optically thin region in a diluted cloud, where particles can stay isolated for a prolonged time. If particles grow, they will initially grow to fractal-like structures, which changes neither their aerodynamic properties nor their optical properties much (\cite{wurm1998}, \cite{martin1986}). In this way, the particles are the perfect sample to be transported to the surface of the protoplanetary disks, i.e., by turbulence. The overall rate of the particle production is an interplay of the attenuation of the incident light flux induced by the cloud of ejected particles and the transportation of the particles by gas drag away from the host body. A prerequisite for this scenario though is that light-induced eruptions actually produce a fraction of small grains. To show this, we carried out laboratory experiments with a basalt dust sample which contains small grains.

\section{Laboratory experiments on small particle generation}
\begin{figure}
\epsscale{.80}
\plotone{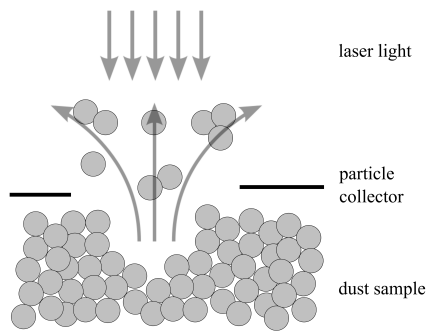}
\caption{Sketch of the experimental setup. A laser illuminates a spot of a dust sample within a vacuum chamber with pressures of some mbar. From the spot particles are released by photophoresis. The released particles are caught by a particle collector and then analyzed by microscopy.}
\label{fig:setup}
\end{figure}

\begin{figure}
\epsscale{.80}
\plotone{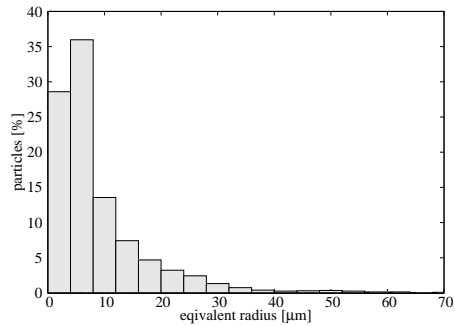}
\caption{Size distribution of the initial basalt powder. The equivalent radius of the grain components is shown. The bin width is 4 $\mu$m and the resolution limit is about $2$ $\mu$m (equivalent radius).}
\label{fig:histo_original}
\end{figure}

\begin{figure}
\epsscale{.80}
\plotone{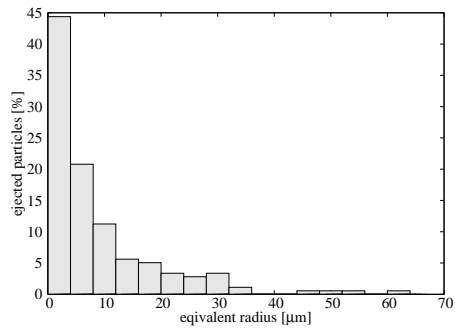}
\caption{Size of photophoretic ejecta. Depicted are the equivalent radii of the ejected particles. The bin width is 4 $\mu$m and the resolution limit is about $2$ $\mu$m (equivalent radius). The light induced erosion process obviously produces small particles.}
\label{fig:histo}
\end{figure}

We carried out photophoretic experiments according to Figures \ref{fig:photopho} and \ref{fig:setup} where we placed a particle collector around the illuminated spot.
As a light source, we used a red laser diode of about 40 mW focused on a spot of $\sim$1 mm$^2$ from above the dust sample. This way the incident light flux is on the order of some $10^4$ W m$^{-2}$ which we regard as appropriate for dust eruptions in the inner region of protoplanetary disks. We used basalt powder as a dust sample. The size distribution of this sample was determined by microscopy and is shown in Figure \ref{fig:histo_original}.
The size distribution of the sampled particles after the light induced eruptions was also analyzed by optical microscopy and the resultant size distribution is plotted in Figure \ref{fig:histo}.
The resolution limit is $\sim$2 $\mu$m. However, the plot shows that the light-induced erosion process produces a large fraction of small particles.

\section{Conclusion}
In this paper, we outlined a possible scenario from which small particles present in protoplanetary disks over millions of years might originate. Our laboratory experiments prove that light-induced eruptions can disassemble a dusty body to its constituents at least down to $\sim 2$ $\mu$m in size. As outlined above, photophoretic sorting will preferentially select the small (sub) micron grains to stay in the optical thin region from where they can be transported to the surface without re-growing to large compact aggregates too rapidly. We conclude that light induced eruptions should be considered as one preferred candidate for the continuous supply of small dust seen at the surface of protoplanetary disks.

\acknowledgments

This work was supported by the DFG.


\begin{thebibliography}{}
\bibitem[Acke \& van de Ancker(2004)]{acke2004} Acke, B., \& van den Ancker, M.E. 2004, Astronomy and Astrophysics, 426, 151
\bibitem[Blum et al.(1996)]{blum1996} Blum, J., Wurm, G., Kempf, S., \& Henning, T. 1996, Icarus, 124, 441
\bibitem[Beresnev et al.(1993)]{beresnev1993} Beresnev, S., Chernyak, V., \& Fomyagin, G. 1993, Physics of Fluids, 5, 2043
\bibitem[Dohnanyi(1969)]{dohnany1969} Dohnanyi, J. S. 1969, Journal of Geophysical Research, 74, 2531
\bibitem[Dullemond \& Dominik(2005)]{dullemond2005} Dullemond, C. P., \& Dominik, C. 2005, Astronomy and Astrophysics, 434, 971
\bibitem[Haisch et al.(2001)]{haisch2001} Haisch, Jr., K. E., Lada, E. A., \& Lada, C. J. 2001, The Astrophysical Journal, 553, L153
\bibitem[Hettner(1924)]{hettner1924} Hettner, G. 1924, Zeitschrift f\"ur Physik, 27, 12
\bibitem[Kaufmann et al.(2007)]{kaufmann2007} Kaufmann, E., K\"omle, N. I., \& Kargl, G. 2007, Advances in Space Research, 39, 370
\bibitem[Kelling \& Wurm(2009)]{kelling2009} Kelling, T., \& Wurm, G. 2009, Physical Review Letters, 103, 215502
\bibitem[Kelling et al.(2011)]{kelling2011} Kelling, T., Wurm, G., Kocifaj, M., Kla\v{c}ka, J., \& Reiss, D. 2011, Icarus, 212, 935
\bibitem[Knudsen(1909)]{knudsen09} Knudsen, M. 1909, Annalen der Physik, 336, 633
\bibitem[Kocifaj et al.(2011)]{kocifaj2011} Kocifaj, M., Kla\v{c}ka, J., Kelling, T., \& Wurm, G. 2011, Icarus, 211, 832
\bibitem[Kocifaj et al.(2010)]{kocifaj2010} Kocifaj, M., Kla\v{c}ka, J., Wurm, G., Kelling, T., \& Koh\'{u}t, I. 2010, Monthly Notices of the Royal Astronomical Society, 404, 1512
\bibitem[Martin(1986)]{martin1986} Martin, J. E., Schaefer, D. W., \& Hurd, A. J. 1986, Phys. Rev. A, 33, 3540
\bibitem[Mathis et al.(1977)]{mathis1977} Mathis, J. S., Rumpl, W., \& Nordsieck, K. H. 1977, The Astrophysical Journal, 217, 425
\bibitem[Meeus et al.(2001)]{meeus2001} Meeus, G., Waters, L.B.F.M., Bouwman, J., van den Ancker, M.E., Waelkens, C., \& Malfait, K. 2004, Astronomy and Astrophysics, 365, 476
\bibitem[Muntz et al.(2002)]{muntz2002} Muntz, E.P., Sone, Y., Aoki, K., Vargo, S., \& Young, M. 2002, Journal of Vacuum Science Technology, 20, 214
\bibitem[Niederd\"orfer(1933)]{niederdoerfer1933} Niederdorfer, E. 1933, Meteorologische Zeitschrift, 50, 201
\bibitem[Okuzumi(2009)]{okuzumi2009} Okuzumi, S. 2009, The Astrophysical Journal, 698, 1122
\bibitem[Olofsson et al.(2010)]{olofsson2010} Olofsson, J., Augereau, J., van Dishoeck, E. F., Mer\'{i}n, B., Grosso, N., M\'{e}nard, F., Blake, G. A., \& Monin, J. 2010, Astronomy and Astrophysics, 520, A39+
\bibitem[Paraskov et al.(2006)]{paraskov2006} Paraskov, G. B., Wurm, G., \& Krauss, O. 2006, The Astrophysical Journal, 648, 1219
\bibitem[Rohatschek(1995)]{rohatschek1995} Rohatschek, H. 1995, Journal of Aerosol Science, 26, 717
\bibitem[Schr\"{a}pler(2011)]{schraepler2011} Schr\"{a}pler, R., \& Blum, J. 2011, The Astrophysical Journal, submitted
\bibitem[Teiser \& Wurm(2009)]{teiser2009} Teiser, J., \& Wurm, G. 2009, Astronomy and Astrophysics, 505, 351
\bibitem[Wood(2000)]{wood2000} Wood, J. A. 2000, Space Science Reviews, 92, 87
\bibitem[Wurm(2007)]{wurm2007} Wurm, G. 2007, Monthly Notices of the Royal Astronomical Society, 380, 683
\bibitem[Wurm \& Blum(1998)]{wurm1998} Wurm, G., \& Blum, J. 1998, Icarus, 132, 125
\bibitem[Wurm \& Krauss(2006)]{wurm2006} Wurm, G., \& Krauss, O. 2006, Physical Review Letters, 96, 134301
\bibitem[Wurm et al.(2005)]{wurm2005} Wurm, G., Paraskov, G., \& Krauss, O. 2005, Icarus, 178, 253
\bibitem[Wurm et al.(2008)]{wurm2008} Wurm, G., Teiser, J., \& Reiss, D. 2008, Geophysical Research Letters, 1, 1
\bibitem[Zolensky(2006)]{zolensky2006} Zolensky, M. E., et al. 2006, Science, 314, 1735
\end{thebibliography}
\end{document}